\input{aipcheck}

\documentclass[
    ,final            
  ]
  {aipproc}

\layoutstyle{6x9}

\begin{document}

\title{Unidentified Galactic High-Energy Sources as Ancient Pulsar Wind Nebulae in the light of new high energy observations and the new code}


\author{O. Tibolla}{
  address={Institut f\"ur Theoretische Physik und Astrophysik, Universit\"at W\"urzburg, Campus Hubland Nord, Emil-Fischer-Str. 31, D-97074 W\"urzburg, Germany}}

\author{M. Vorster}{
  address={Center for Space Research, North-West University, Potchefstroom, South Africa}
}

\author{O. de Jager}{
  address={Center for Space Research, North-West University, Potchefstroom, South Africa}
}

\author{S.E.S. Ferreira}{
  address={Center for Space Research, North-West University, Potchefstroom, South Africa}
}

\author{S. Kaufmann}{
  address={Landessternwarte, Universit\"at Heidelberg, K\"onigstuhl 12, D-69117 Heidelberg, Germany}
}

\author{C. Venter}{
  address={Center for Space Research, North-West University, Potchefstroom, South Africa}
}

\author{K. Mannheim}{
  address={Institut f\"ur Theoretische Physik und Astrophysik, Universit\"at W\"urzburg, Campus Hubland Nord, Emil-Fischer-Str. 31, D-97074 W\"urzburg, Germany}
}

\author{F. Giordano}{
  address={INFN \& Dipartimento di Fisica ``M. Merlin'', Universit\`a e Politecnico di Bari, I-70126 Bari, Italy}
}

\begin{abstract}
 
In a Pulsar Wind Nebula (PWN), the lifetime of inverse Compton (IC) emitting e$^-$ exceeds the lifetime of its progenitor pulsar (as well as its shell-type remnant), but it also exceeds the age of those that emit via synchrotron radiation. Therefore, during its evolution, the PWN can remain bright in IC so that its GeV-TeV gamma-ray flux remains high for timescales much larger (for $10^5-10^6$ yrs) than the pulsar lifetime and the X-ray PWN lifetime. In this scenario, the magnetic field in the cavity induced by the wind of the progenitor star plays a crucial role. This scenario is in line with the discovery of several unidentified or ``dark'' sources in the TeV gamma-ray band without X-ray counterparts; and it is also finding confirmation in the recent discoveries at GeV gamma rays. Moreover, these consequences could be also important for reinterpreting the detection of starburst galaxies in the TeV gamma-ray band when considering a leptonic origin of the gamma-ray signal. Both theoretical aspects and their observational proofs will be discussed, as well as the first results of our new modeling code.

\end{abstract}

\maketitle


\section{Ancient Pulsar Wind Nebulae}

Although they are often detected as non-thermal X-ray sources, evolved PWNe can indeed lead to a bright $\gamma$-ray source without any counterpart \cite{1} \cite{2} \cite{3}. The key issue is that the low-energy synchrotron emission,  depends on the internal PWN magnetic field \cite{4} which may vary as a function of time, following $\tau_E \propto t^{2 \alpha}$ if $B(t) \propto t^{-\alpha}$, where $\tau_E$ is the synchrotron emitting lifetime of TeV $\gamma$-ray emitting leptons with energy E and $\alpha$ is the power-law index of the decay of the average nebular field strength, whereas the VHE emission depends on the background radiation fields, which are constant on time-scales relevant for PWN evolution. MHD simulations of composite SNRs \cite{5} find that $\alpha = 1.3$ until the passage of the reverse shock (RS); after the field decay should continue for much longer time since expansion continues.
From hydrodynamic simulations, an expression for the time of the RS passage has been given \cite{6} as $T_R = 10 \mathrm{kyr} (\frac{\rho_{ISM}}{10^{-24} \mathrm{g}/\mathrm{cm}^{-3}})^{-\frac{1}{3}} (\frac{M_{ej}}{10 M_{Sun}})^{\frac{3}{4}} (\frac{E_{ej}}{10^{52} \mathrm{erg}})^{-\frac{2}{3}}$, where the $1^{st}$ term is the density of the ISM, the $2^{nd}$ the ejecta mass during the SN explosion and the $3^{rd}$ the SN blast wave energy. The stellar wind of a high-mass star can blow a cavity around the progenitor star \cite{cav} with relatively low ISM density, so that $T_R >> 10$ kyr. In such a case B(t) can decay as $t^{-\alpha}$, until the field is low enough for the X-ray flux to drop below the typical sensitivity levels. As a result, in a scenario where the magnetic field decays as a function of time, the synchrotron emission will also fade as the PWN evolves. The reduced synchrotron losses for high-energy e$^-$ for such a scenario will then lead to
increased lifetimes for these leptonic particles. For timescales shorter than the inverse-Compton lifetime of the e$^-$ ( $t_{IC} \propto 1.2 \times 10^6 (E_e/1 \mathrm{TeV})^{-1}$ years), this will result in an accumulation of VHE e$^-$ (e.g. seen in HESS J1825-137)  which will also lead to an increased $\gamma$-ray production via IC. To summarize, during their evolution PWN may appear as $\gamma$-ray sources with only very faint low-energy counterparts and this may represent a viable model for many unidentified TeV sources. This effect can be clearly modeled, tested on young PWNe such as the Crab or G21.5-0.9, and finally applied to some candidate ancient PWNe such as HESS J1427-608 and HESS J1507-622.

\vspace{-0.5cm}
\section{The new code}

In this code (work in progress) we adopted the time evolution of the lepton spectrum of \cite{8} (eq. 7 and 11), 
and the source function for the system $Q(E_e,t)$ given by \cite{9}

\begin{displaymath}
Q(E_e,t) = \left\{ \begin{array}{ll}
Q_{\rm{r}}(t)\left(E_e/E_b\right)^{-p_1} & \textrm{if $E_{\min}<E_e<E_{\rm{b}}$}\\
Q_{\rm{x}}(t)\left(E_e/E_b\right)^{-p_2} & \textrm{if $E_{\rm{b}}<E_e<E_{\max}$} 
\end{array} \right.
\end{displaymath}

where the normalisation factors $Q_i(t)$ are determined using the prescription that the energy in the lepton spectrum should be some fraction of the spin-down luminosity $L(t)$:
$\int Q_i(E_e,t)E_e dE_e = \eta_iL(t)$. Note that the two normalisation factors are solved independently, with the only requirement that $\eta_{\rm{r}}+\eta_{\rm{x}} < 1$.  The time evolution of the spin-down luminosity is given by $L(t) = \frac{L_0}{(1+t/\tau)^2}$, where $\tau$ is the characteristic spin-down time-scale of the pulsar.  The synchrotron $\tau_{\rm{syn}}$ and escape $\tau_{\rm{esc}}$ time-scales can respectively be approximated by \cite{8} $\tau_{\rm{syn}} (t) \approx 1.25 \times 10^7 \left(\frac{1 \mbox{ $\mu$G}}{B(t)}\right)^2 \left(\frac{1 \mbox{ TeV}}{E_e}\right) \mbox{ yr}$ and $\tau_{\rm{esc}} (t) \approx 3.4 \times 10^4 \left(\frac{B(t)}{1 \mbox{ $\mu$G}}\right) \left(\frac{1 \mbox{ TeV}}{E_e}\right) \left(\frac{ r_{\rm{pwn}}(t)}{1 \mbox{ pc}}\right)^2 \mbox{ yr}$.

Since $\alpha = 1.3$ until the passage of the RS, the evolution of the radius of the PWN can be approximated by

\begin{displaymath}
r_{\rm{pwn}}(t) = \left\{ \begin{array}{ll}
r_0(t/t_0)^{1.4} & \textrm{if $t<t_{\rm{rs}}$}\\
r_0(t/t_0)^{0.2} & \textrm{if $t>t_{\rm{rs}}$}
\end{array}  \right.
\end{displaymath}

\begin{figure}
  \includegraphics[width=0.42\columnwidth]{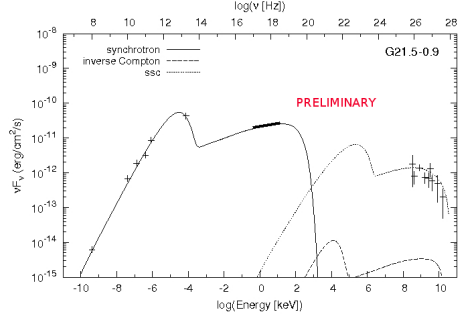}
  \caption{The SED of G21.5-0.9 shows that our new model can describe well the spectral features of young PWNe. The data are taken from \cite{3}.}
\vspace{-2cm}
\end{figure}

where $r_0$ is a normalisation constant and $t_{\rm{rs}}=5000$ yr the time needed for the RS of the supernova shell remnant to reach the PWN.  For the PWN radius we assume a power-law expansion with index 1.4 before the passage of the RS and index 0.2 after it.


\subsection{Testing the new code on young PWNe (G21.5-0.9)}

In order to test the consistency of our code we tested it on the Crab nebula and on G21.5-0.9, obtaining
compatible results with \cite{8} and with its previous version \cite{1} \cite{3}. The derived value for the
magnetic field of G21.5-0.9 is similar to the value estimated for the 950 yr old Crab Nebula. Other
parameters derived are listed in Tab. 1. The results, shown in Fig. 1 and summarized in Tab. 1, demonstrate that the
new code is describing well the spectral energy distribution (SED) of G21.5-0.9. 

In order to fit the radio and X-ray synchrotron emission, it is necessary for
the e$⁻$ spectrum to have a discontinuity between the two parts of the broken
power-law. A similar feature has been discussed by \cite{10} to account for the emission
from the Vela PWN. Despite the different parent e$^-$ spectrum, the results are compatible with \cite{3}.

\vspace{-0.5cm}
\subsection{The code describes well also Unidentified sources}

\begin{figure}
\hbox{
\includegraphics[width=0.4\textwidth]{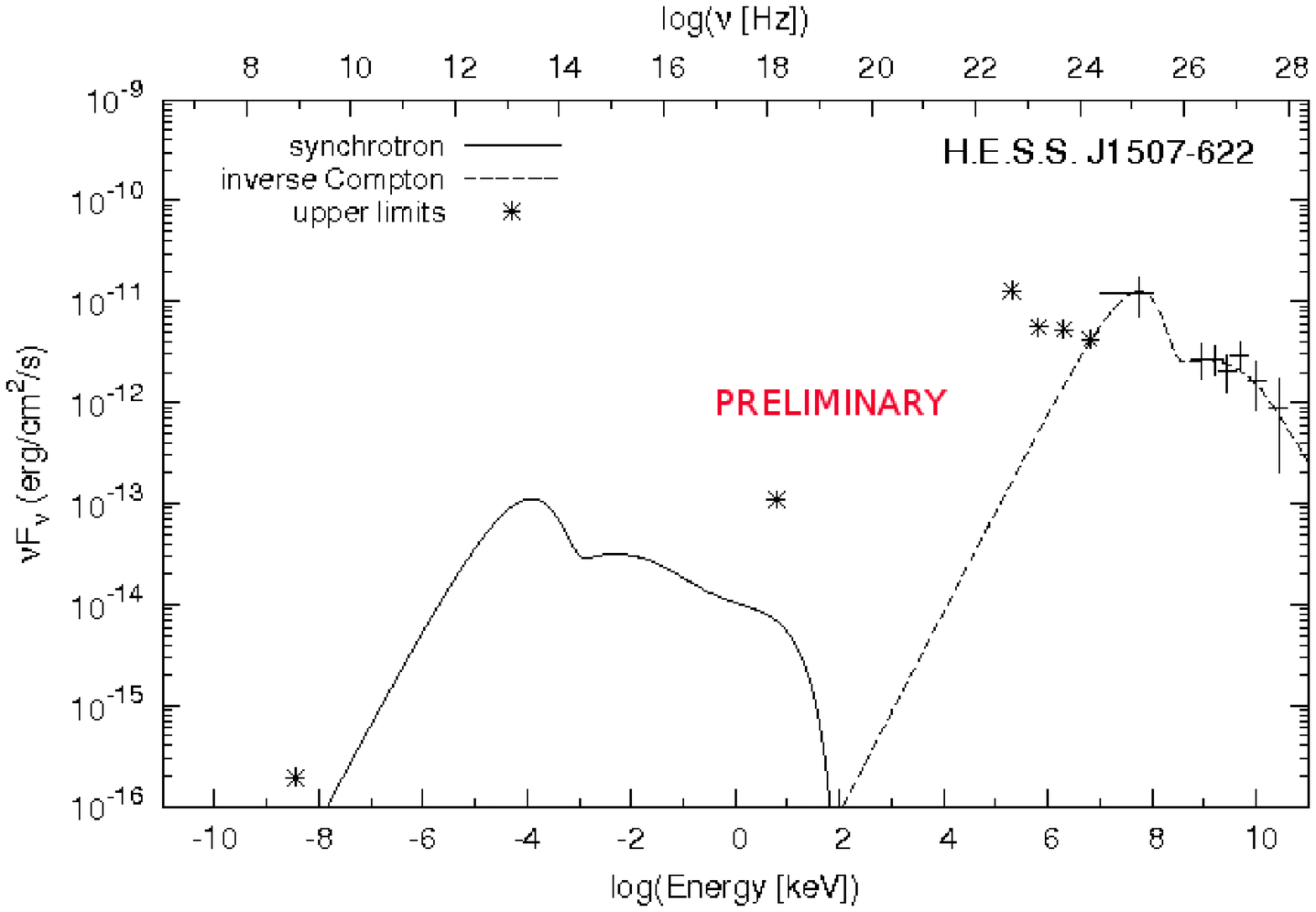}
\hspace{1.0cm}
\includegraphics[width=0.4\textwidth]{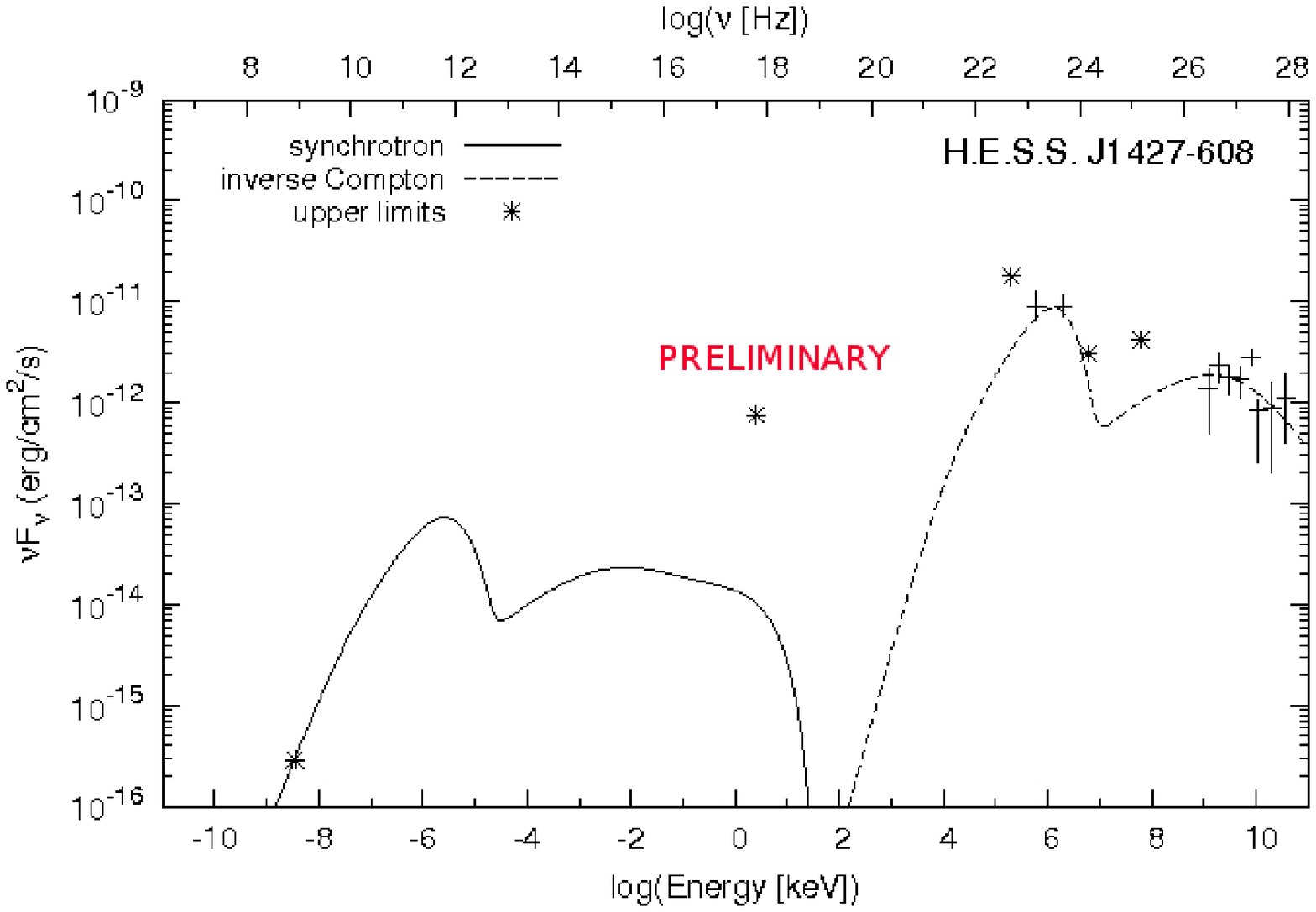} 
}
\caption{\footnotesize {
{\it left:} SED of HESS J1507-622 modeled with the new code; the radio information is taken from \cite{23}, the \emph{Fermi}-LAT points from \cite{11}, the Chandra and H.E.S.S. points from \cite{2}.
{\it right}: SED of HESS J1427-608 modeled with the new code; data are taken from \cite{uni}.
}
}
\end{figure}
\vspace{-0.5cm}
After testing the validity of the code on young, well-known PWNe, we test it on Unidentified high-energy $\gamma$-ray sources and especially on the so called ``dark sources'', i.e. on the $\gamma$-ray sources that, even after deep multi-wavelength observations, lack any plausible X-ray counterparts (such as HESS J1507-622, HESS J1427-608, HESS J1708-410 and HESS J1616-508).
Here we show the results on HESS J1507-622 and HESS J1427-608. The results are summarized in Fig. 2 and in Tab. 1, showing that our new code describes the SEDs also of these 2 ``dark sources'' very well, strengthening the idea that we are dealing with ancient PWNe.

It is important to underline that for HESS J1507-622 (the best example
for this kind of sources \cite{2} and the most challenging for the model, especially given the very
low radio upper limit) the age (15 kys) is perfectly compatible with the estimation derived in  \cite{3}, as is the distance (6 kpc) with the lower limit of \cite{2}. Moreover, since there is no pulsar detected in the region of HESS J1507-622 or HESS J1427-608
(as expected for such old PWNe that the pulsar could have
already spun down below the sensitivity of the current instruments \cite{5}), an arbitrary
choice of the initial conditions for the pulsar parameters is required; moreover, in order to fit the data for HESS J1507-622, we need a current
magnetic field of about 0.3 $\mu$G. This value is compatible with the upper limit of 0.5 $\mu$G derived in \cite{2}, however we should underline that this value is smaller than the ISM magnetic field (that would raise again the question of section 3 of \cite{3}).

\vspace{-0.5cm}
\section{Summary and conclusions}

\begin{table}
\begin{tabular}{cccc}
\hline
   \tablehead{1}{r}{b}{~~~}
  & \tablehead{1}{r}{b}{G21.5-0.9}
  & \tablehead{1}{r}{b}{HESS J1507-622}
  & \tablehead{1}{r}{b}{HESS J1427-608}   \\
\hline

$L_0$ ($10^{37}$ erg/s) &  $5$ & $100$ & $100$ \\
$T_{\rm{age}}$ (yr) & $870$ & $15000$ & $15000$\\
Distance (pc) & $4400$ & $6000$ & $6000$ \\
$\tau$ (yr) & $4800$ & $6000$ & $6000$\\
$B(T_{\rm{age}})$ ($\mu$G) & $250$ & $0.3$ & $0.3$\\
$\eta_{\rm{r}}$ & $0.3$ & $0.2$ & $0.7$ \\
$\eta_{\rm{x}}$ & $0.06$ & $0.05$ & $0.05$ \\
$E_{\rm{max}}$ (TeV) & $10^{2}$ & $10^{3}$ & $6 \times 10^{2}$\\
$E_{\rm{min}}$ (TeV) & $10^{-4}$ & $10^{-3}$ & $5 \times 10^{-2}$\\
$E_{\rm{b}}$ (TeV) & $0.06$ & $4$ & $0.6$\\

\hline
\end{tabular}
\caption{Model parameters for the three sources discussed in the text.}
\label{tab:a}
\end{table}

Our new code represents an useful tool to describe PWNe, since it could describe so far both
their young and aged phases. The preliminary results on the three sources shown here are summarized in Tab. 1.
Together with the new high-energy observations \cite{11}, it is strengthening the idea that VHE unidentified sources can indeed explained as ancient
PWNe. Regarding the used model, we notice that: (1) the e$^-$ energy spectrum needs to be described by a discontinuous broken power-law in order to
simultaneously fit all the MWL data; and (2) the fraction of spin-down power needed for the higher-energy
part of the e$^-$ spectrum is found to be $\sim$10 times smaller than the energy needed for the low-energy part of the spectrum.
The consequences on Galactic and extragalactic (e.g. Starburst galaxies \cite{sb}) astrophysics, and consequently on CRs studies, are remarkable and
already discussed in \cite{2} \cite{3} \cite{5}.

\bibliographystyle{aipproc}

\end{document}